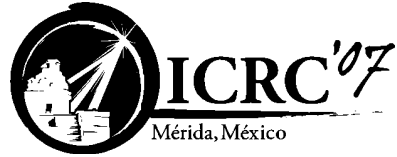

# Simulation of Cosmic Ray propagation in the Galactic Centre Ridge in Accordance with Observed VHE γ-ray Emission

DIMITRAKOUDIS STAVROS[1], MASTICHIADIS APOSTOLOS[1], GERANIOS ATHANASIOS[2]

[1]*University of Athens, Physics Department, Section of Astrophysics, Astronomy and Mechanics, Panepistimioupoli 15771, Greece*

[2]*University of Athens, Nuclear and Particle Physics Department, Panepistimioupoli 15771, Greece*

steeve_dim@hotmail.com

**Abstract:** Diffuse VHE γ radiation from the Galactic Centre ridge observed by the H.E.S.S. telescope has been convincingly linked with the propagation of recently accelerated cosmic rays that interact with molecular hydrogen clouds during their diffusion. Through a series of time-dependent simulations of that diffusion for different propagation parameters we have obtained the most probable values of the diffusion coefficient for the Galactic Centre region. Assuming that the diffusion coefficient is of the form $\kappa(E) = \kappa_0 (E/E_0)^\delta$, then for different optimal combinations of $\kappa_0$ and $\delta$ its value is obtained for cosmic rays originating from a central point (possibly Sgr A East) 10 kyr ago.

## Introduction

Since 2004 the High Energy Stereoscopic System (H.E.S.S.) has provided us with images of VHE γ-ray emission from the Galactic Centre of unprecedented angular resolution [5]. Besides the discovery of a few point sources of γ-rays, the H.E.S.S. collaboration presented a γ-ray count map of the inner 400pc of the galaxy that shows lower intensity emission distributed over an area that spans approximately 300pc [1]. This emission seems to follow the molecular gas density as measured by its CS distribution [13], up to a distance of approximately 1.3° in galactic longitude from the Galactic Centre.

Such a correlation hints at the source of the γ-rays, for which two theories have been proposed. One possibility is that a population of electron accelerators produces the observed emission via inverse Compton scattering. The objects that would make up such a population, such as pulsar wind nebulae, would thrive in regions of high-density molecular gas, but the large number of such sources needed to reproduce the observed emission renders this possibility unlikely. The other possibility is that the γ-rays are produced when cosmic rays interact with the molecular gas, producing pions that decay into photons.

The lower energy threshold of the H.E.S.S. survey was 380 GeV, so cosmic ray protons of higher energies would be needed to produce the observed γ-rays. Furthermore, those cosmic ray protons would have to have been accelerated near the Galactic Centre at some point in the past, yet not diffused significantly beyond 1.3° from it. Assuming the validity of this theory, we may use this data to infer the diffusion properties of cosmic rays in the Galactic Centre region for possible sources of cosmic rays.

Aharonian et al. [1] already proposed the supernova remnant Sgr A East, with an estimated age of 10 kyr, or the putative supermassive black hole Sgr A* with a more remote age of activity, as possible sources. Büsching et al. [3] studied the first possibility in an analytical reproduction of the H.E.S.S. observations by calculating the emission results for different diffusion coefficients. Using a small number of Gaussian functions to represent the molecular clouds and protons of mean energy ~ 3 TeV to represent the cosmic rays they arrived at a diffusion coefficient of  $\kappa = 1.3$ kpcMyr$^{-1}$.

In our paper we used a series of time-dependent simulations of proton propagation in the Galactic Centre for different diffusion parameters, in a 3D environment comprising of over 500 distinct molecular clouds synthesized from a variety of observations. Through this process we have arrived at a new estimation of the diffusion coefficient in the Galactic Centre region.

## Synthesizing a hydrogen cloud map

The environment in which we simulated the diffusion of cosmic rays is an area rich in $H_2$ gas, contained in a complex setup of high-density clouds, ridges and streams comprising about 10% of our galaxy's interstellar molecular gas, i.e. about 2 to 5 x $10^7$ solar masses [13]. Due to the high densities involved, tracer molecules are used to determine the mass of each gas cloud. To create a realistic 3D map of that environment we first obtained the data for the 159 distinct molecular cloud clumps recognised by Miyazaki & Tsuboi [9], i.e. their assigned galactic longitudes, latitudes, radii and densities. Their radial distances are unknown, so we assumed a random function to simulate them. To this data we added the locations and densities of the radio-sources Sgr A [11], Sgr B1, Sgr B2 and Sgr C [6], which are rich in atomic hydrogen. We then broke down the entire CS map of the Galactic Centre by Tsuboi et al. [13] into blocks of uniform density, which we treated as clouds of equal radii with random radial distances. Finally we added the larger CO clouds by Oka et al. [10] for an expanded distribution in our map. Every time we added a new set of clouds we checked the possibility that clouds from previous sets were sharing their projected locations with new clouds, and we subtracted the masses of such clouds accordingly. The result was 584 clouds of hydrogen gas with a high uncertainty as to their radial distances. These clouds were then turned into a 3D grid of 120x60x60 boxes of uniform density, which form the volume of our diffusion model.

## Diffusion model

We used the diffusive model of CR propagation, assuming a diffusion coefficient of the form:

$$\kappa = \kappa_0 (E/E_0)^\delta$$

where E is the energy of the CR protons, $E_0$=1GeV, while $\kappa_0$ and $\delta$ are free parameters, whose values we will try to infer through our simulations. The index $\delta$ is a measure of the turbulence of the magnetic field in the GC region, and is assumed to be $0.3 < \delta < 0.6$ [12], while $\kappa_0$ is a constant whose value we will seek to ascertain for each value of $\delta$.

The diffusion coefficient in our simulations is represented by the mean free path $\ell$

$$\ell = 3\kappa/c.$$

Thus, our test protons move in straight lines of length equal to $\ell$, after which their directions change randomly. At the end of each such free walk, the box number of the hydrogen density grid is calculated and a check is made for collisions with hydrogen protons. If such a collision occurs, the diffusion continues with a lower proton energy, as shown in the next section.

## Production of γ-rays

γ-rays are produced from neutral pion decay. Pions are produced in proton-proton collisions, with two main distinct possibilities

a) $p + p \rightarrow p + p + \pi^+ + \pi^- + \pi^0$
b) $p + p \rightarrow n + p + \pi^+$

In the first case the initial proton loses part of its energy and a multiplicity of pions is created that share equally the energy lost from the proton. The positive and negative pions break down into electrons and neutrinos, while the neutral pions produce γ-rays. Assuming an inelasticity $k_{pp}$=0.45 [8], 15% of the initial proton energy will go to the produced γ-rays, while the initial proton will continue its diffusion with 55% of its initial energy.

We assume that the cross section for this reaction between protons is $\sigma_{pp} = 4 \cdot 10^{-26}$ cm$^2$ [2], while an increase by a factor of 1.30 is needed to account for the known chemical composition of the Interstellar Medium [7].

In the second case no photons are produced, but the initial proton is turned into a neutron. It will thus continue its propagation in a straight line, unaffected by the magnetic fields that regulated its trajectory as a proton. However it will revert back into a proton with a half life $\tau = 886.7 \pm 0.8$ sec [15]. Accounting for time dilation, the distances traveled as a neutron are always much smaller than the mean free paths, so we can easily assume that this case will not affect the diffusion process, besides the reduction in proton energy.

In each case, a reaction probability is calculated for each step in the random walk. The cross section defines a cylinder along the mean free path that contains a density n of hydrogen molecules, that is retrieved from the grid box where each step ends. The reaction probability is RPU = $n\ell\sigma_{pp}$ and it is checked against a random number at the conclusion of each step. If the random number is smaller than RPU, then there is a collision and γ-rays are produced.

## Simulation parameters

If we assume a single source for the diffuse cosmic rays in the Galactic Centre, then that would be either the supernova remnant Sgr A East or the putative supermassive black hole Sgr A*. The former has a calculated age of $10^4$ yr [14] while the latter could have had a burst of activity even further in the past. In our simulations we have focused on an age of $10^4$ yr and a source at galactic coordinates l = 0°, b = 0°, which corresponds to both candidates. We have assumed a production time for the cosmic rays of $10^2$ yr and they are broken down into six energetic populations, ranging from E = $10^{12.5}$ eV to E = $10^{15}$ eV. Each population has the same number of test particles, but those numbers are normalised after each simulation according to a power law distribution that fits best the power law of the observed γ-rays from H.E.S.S. The resulting γ-rays are then compared against the results from H.E.S.S. using the reduced $\chi^2$ criterion.

## Results and discussion

The resulting reduced $\chi^2$ values for different diffusion coefficients are shown in Fig. 1. The minimum value of $\chi^2$ observed is 1.69, for δ = 0.4 and $\kappa_0$ = 0.126 kpc$^2$Myr$^{-1}$, but for all values of δ we can see minima of $\chi^2$ that are at most equal to 2. If we use those values of δ and $\kappa_0$ to calculate the diffusion coefficient for protons of energy $10^{12.5}$eV (the lowest energy in our sample and also the most important due to the relative abundance of such protons over those of higher energies), we will arrive at the results illustrated in Fig. 2. There we can see that for each value of δ the diffusion coefficient displays the same minimum at κ ≃ 3 kpc$^2$Myr$^{-1}$. This value is higher than that calculated by Büsching et al. [3] but close to that suggested by Aharonian et al. [1] (less than 3.5 kpc$^2$ Myr$^{-1}$).

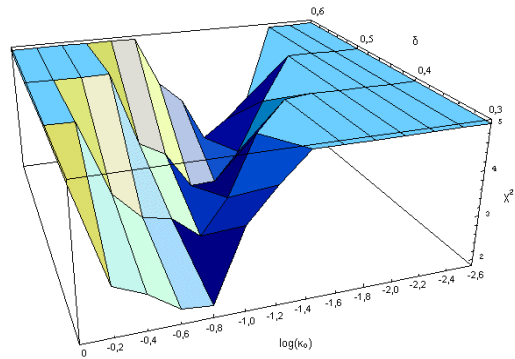

Figure 1: Reduced $\chi^2$ values for different values of $\kappa_0$ and δ.

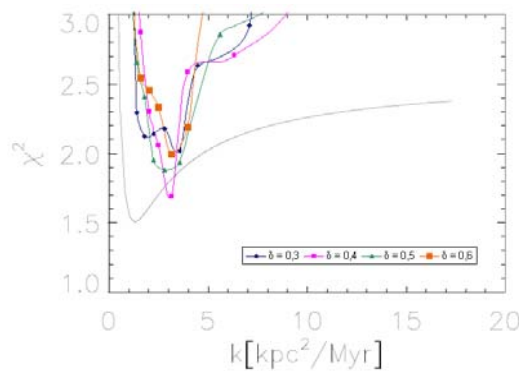

Figure 2: The colored curves represent the reduced $\chi^2$ values for different diffusion coefficients, for each value of δ. We see in all cases a minimum for κ ≃ 3 kpcMyr$^{-1}$. The grey line shows the equivalent results from Büsching et al. [3] for comparison.

These results were derived under the assumption that the acceleration of the cosmic rays responsible for the observed γ-rays occurred $10^4$ yrs ago at a single source, with no subsequent periods of activity at that source. Recent papers [4] have noted that this may be a very simplified approach, as there have been many supernovae in the Galactic Centre region in the past millennia. Also the uncertainty in the radial distances of hydrogen concentrations may have had a significant impact on the final results. Finally, these simulations assumed that the diffusion coefficient remains constant throughout the whole region of propagation, and local orderings of magnetic fields were not taken into account.

Those limitations notwithstanding, the results of our simulations are useful in providing an estimate of the diffusion coefficient in the Galactic Centre, taking the different magnetic turbulence theories (that become manifest in the different values of δ) into account.

## Acknowledgments

This project is co-funded by the European Social Fund and National Resources – (EPEAEK II) PYTHAGORAS II.

## References

[1] Aharonian, F.A. and 101 co-authors (H.E.S.S. Collaboration). Discovery of very-high-energy γ-rays from the Galactic Centre ridge, Nature, 439,695, 2006.
[2] Begelman, M.C., Rudak, B., Sikora, M. Consequences of Relativistic Proton Injection in Active Galactic Nuclei, ApJ, 362, 38, ApJ, 362, 38, 1990.
[3] Büsching, I., de Jager, O.C. and Snyman, J. Obtaining cosmic ray propagation parameters from diffuse VHE γ-ray emission from the Galactic center ridge, arXiv:astro-ph/0602193, 2006.
[4] Erlykin, A.D. and Wolfendale, A.W. Gamma Rays from the Galactic Centre, arXiv: astro-ph/0705.2333v1, 2007.
[5] Hinton, J. (H.E.S.S. Collaboration). The H.E.S.S. View of the Central 200 Parsecs, arXiv:astro-ph/0607351, 2006.
[6] Law, C. and Yusef-Zadeth, F. Proceedings of X-Ray and Radio Connections Santa Fe, New Mexico, 3-6 February 2004.
[7] Mannheim, K. and Schlickeiser, R. Interactions of Cosmic Ray Nuclei, A&A, 286, 983, 1994 .
[8] Mastichiadis, A., Kirk, J.G. Self-consistent particle acceleration in active galactic nuclei, A&A, 295, 613, 1995.
[9] Miyazaki, A. and Tsuboi, K. Dense molecular clouds in the Galactic Center region. II. Statistical properties of the Galactic Center molecular clouds, ApJ, 536, 357, 2000.
[10] Oka, T., Hasegawa, T., Hayashi, M., Handa, T. and Sakamoto, S. CO (J = 2-1) line observations of the Galactic Center molecular cloud complex. II. Dynamical structure and physical conditions, ApJ, 493, 730, 1998.
[11] Shukla, H., Yun, M. S., Scoville, N. Z. Dense Ionized and Neutral Gas Surrounding Sgr A*, ApJ, 616, 231, 2004.
[12] Strong, A.W., Moskalenko, I.V., Ptuskin, V.S. Cosmic-ray propagation and interactions in the galaxy, submitted for publication to the Annual Review of Nuclear and Particle Science (v.57), arXiv:astro-ph/0701517, 2007.
[13] Tsuboi, K., Toshihiro, H., & Ukita, N. Dense molecular clouds in the Galactic Center region. I. Observations and data, ApJS, 120, 1, 1999.
[14] Uchida K.I., Morris M., Serabyn E., Fong D., Meseroll T. in Sofue Y., ed, Proc. IAU Symp. 184, The Central Regions of the Galaxy and Galaxies, Kluwer, Dordrecht, p.317, 1998.
[15] Yao, W.M. and co-authors (Particle Data Group). Review of Particle Physics, J. Phys. G 33, 1 (URL: http://pdg.lbl.gov), 2006.